\documentclass[pra,twocolumn,superscriptaddress,nofootinbib,noshowpacs,preprintnumbers,longbibliography,floatfix]{revtex4-2}
\usepackage[utf8]{inputenc}
\usepackage[english]{babel}
\usepackage{graphicx}
\usepackage{subfigure}
\usepackage{float}
\usepackage{amssymb}
\usepackage{amsmath}
\usepackage{dsfont}
\usepackage{array}
\usepackage{bm,fixmath}
\usepackage{mathrsfs}
\usepackage[normalem]{ulem}
\usepackage[usenames, dvipsnames]{xcolor}
\usepackage{physics}
\usepackage{tikz}
\usepackage{qcircuit}
\usepackage{placeins}
\usepackage{subfigure}
\usepackage{verbatim}
\usepackage[pdftex,
            pdftitle={Iterative ITE},
	    pdfauthor={Yahui Chai, Alice Di Tucci},
            bookmarks,
            colorlinks,
            linkcolor=myblue,
            citecolor=mymagenta,
            menucolor=black,
            urlcolor=myblue,
            plainpages=false,
            pdfpagelabels,
            hypertexnames=false]{hyperref}
\usepackage{orcidlink}
\usepackage{comment}
\graphicspath{{figure/}}

\definecolor{mymagenta}{RGB}{200, 0, 100}
\definecolor{myblue}{RGB}{45, 48, 146}

\begin{document}

\title{Optimizing QUBO on a quantum computer by mimicking imaginary time evolution}

\author{Yahui Chai}
\affiliation{Deutsches Elektronen-Synchrotron DESY, 15738 Zeuthen, Germany}

\author{Alice Di Tucci}
\affiliation{Deutsches Elektronen-Synchrotron DESY, 15738 Zeuthen, Germany}
\affiliation{Physikalisch-Technische Bundesanstalt PTB, 38116 Braunschweig, Germany}
\email{alice.ditucci@desy.de}
\date{\today}

\begin{abstract}
\noindent
We propose a hybrid quantum-classical algorithm for solving QUBO problems using an Imaginary Time Evolution-Mimicking Circuit (ITEMC). The circuit parameters are optimized to closely mimic imaginary time evolution, using only single- and two-qubit expectation values. This significantly reduces the measurement overhead by avoiding full energy evaluation. By updating the initial state based on results from last step iteratively, the algorithm quickly converges to the low-energy solutions. With a pre-sorting step that optimizes quantum gate ordering based on QUBO coefficients, the convergence is further improved. Our classical simulations achieve approximation ratios above $99\%$ up to 150 qubits. Furthermore, the linear scaling of entanglement entropy with system size suggests that the circuit is challenging to simulate classically using tensor networks. We also demonstrate hardware runs on IBM's device for 40, 60, and 80 qubits, and obtain solutions compatible with that from simulated annealing.
\end{abstract}

\maketitle


\section{Introduction}
\label{sec:introduction}
Quantum computing has emerged as a promising paradigm for solving complex optimization problems that are intractable for classical methods. One prominent class of such problems is the Quadratic Unconstrained Binary Optimization (QUBO) problem, which is central to many combinatorial optimization tasks, with applications across a wide range of fields, including finance, logistics, machine learning, and material science~\cite{Kochenberger:2014ful}. QUBO problems are generally NP-hard~\cite{Barahona:1982}, making their efficient solution a key challenge in both classical and quantum computing. Several quantum algorithms have been developed to address QUBO problems~\cite{Abbas:2023agz}, with notable approaches including the Variational Quantum Eigensolver (VQE)~\cite{Peruzzo:2013bzg, cerezo2021variational} and the Quantum Approximate Optimization Algorithm (QAOA)~\cite{Farhi:2014ych}. VQE is widely used for solving problems in quantum chemistry and optimization~\cite{Tilly:2021jem, Kandala:2017vok, Lim:2024bvw, Fonseca:2025wkx, Chai:2023ixt, Chai:2023fqb, Amaro_2022, Nannicini_2019, Chai:2024sca, Wang:2024jis}, QAOA and its variants have gained significant attention as a general-purpose algorithm for combinatorial optimization \cite{Blekos:2023nil, Sachdeva:2024kob, He:2023pue, Sack:2023cvk, Herman:2022tto, yu_abQAOA_2025, chai_SQAOA_2022, sachdeva_QAOA127qubit_2024, Chandarana:2021kik, DC_QAOA_2022}. Both methods rely on a hybrid quantum-classical framework, where a quantum computer prepares a trial state and a classical optimizer updates the parameters of the quantum circuit to minimize the cost function. However, these algorithms suffer from practical limitations, such as high measurement costs, large classical optimization overhead, and challenges in scaling to large systems~\cite{Anschuetz_2022, Larocca:2024plh, Schwagerl:2024xqd}.

In this work, we propose a novel hybrid quantum-classical algorithm inspired by Imaginary Time Evolution (ITE), a technique traditionally used in quantum mechanics to find ground states of Hamiltonians. The central idea behind ITE is to simulate the dynamics of a system evolving in imaginary time, which causes high-energy states to decay exponentially faster than the ground state. While the ITE can not be implemented on a quantum computer directly, several previous works have explored variational approaches to approximate ITE~\cite{Motta_2019, McArdle_2019, PRXQuantum.2.010317, Yuan2019}, typically requiring substantial resources to determine circuit parameters. More recent developments have reduced circuit depth and evaluation cost~\cite{Amaro:2021acj, PhysRevResearch.3.033083, Morris:2024yne}, but still target accurate replication of ITE and remain resource-intensive. In contrast, our approach constructs a parameterized quantum circuit that approximately mimics ITE using a sequence of simple, efficiently optimized gate operations. Crucially, the improvement in solution quality does not depend on simulating a large imaginary time but instead arises from iteratively updating the initial state based on previous measurement outcomes. This iterative strategy allows the solution to improve progressively, without increasing circuit depth or the number of variational parameters.

Our approach builds on the work in Ref.~\cite{Chai:2024sca}, in which ITE was used to prepare an initial state that provided a good starting point for the classical optimization of the VQE. Here, we expand on this idea by focusing on the ITE-mimicking part itself, and importantly, we eliminate the need for VQE altogether. By doing so, we reduce the complexity and measurement overhead associated with the classical optimization step. The core of our approach involves two key innovations. First, we design an ITE-mimicking circuit (ITEMC) with parameters determined by small number of shots, and use it to evolve the quantum state iteratively, with the initial state updated base on results from the previous step. Second, we introduce a pre-sorting step that optimizes the ordering of quantum gates based on the QUBO coefficients. These innovations enable the algorithm to efficiently converge to the low-energy solution and with much fewer measurements than VQE.

The paper is organized as follows. Section \ref{sec:qubo} explains the QUBO problem and provides the necessary background. In Section \ref{sec:ITEMC}, we introduce the ITEMC algorithms. Section \ref{sec:results} presents our classic numerical results up to 150 qubits. The results of the hardware runs with 40, 60, and 80 qubits are presented in Section \ref{sec:hardware}. The paper concludes with a discussion of our findings and an outlook on potential avenues for future research in Sec.\ref{sec:summary}.

\section{QUBO}\label{sec:qubo}

QUBO problems are widely used in combinatorial optimization problems, including Graph partitioning~\cite{hayato_2017}, Traveling salesman problems~\cite{Jain_2021}, Flight-gate assignment problems~\cite{Chai:2023ixt, Chai:2023fqb}, among others. The goal is to minimize (or maximize) a quadratic function defined over binary variables $x_i \in \{0, 1\}$:
\begin{equation}
    f(\bold{x}) = \bold{x}^T Q \bold{x} = \sum_{i = 0}^{N-1} \sum_{j=0}^{N-1} x_i \cdot Q_{ij} \cdot x_j.
\end{equation}
where $Q \in \mathbb{R}^{N \times N}$ is a matrix that encodes the problem's interactions and biases, and $\bold{x} = \{x_0, x_1, \cdots x_{N-1}\}$ is a vector of $N$ binary variables.  Each binary configuration defines a different value of the cost function and the task is to find the one that minimizes it.

QUBO problems can be represented as a graph by treating the matrix $Q$ as the adjacency matrix of a weighted, undirected graph $G(V, E)$. The set of vertices $V$ corresponds to the binary variables, with $|V| = N$ representing the total number of variables. The edges $E$ are defined by the non-zero off-diagonal elements of $Q$, meaning there is an edge between vertices $i$ and $j$ if $Q_{ij}=Q_{ji}\neq 0$.  In the following, we consider various graph structures, including 3-regular graphs, complete graphs, and graphs of intermediate density $d = |E| / \binom{|V|}{2} $, with $ \binom{|V|}{2} $ the total number of possible edges in a complete graph. 

In classical computing, solving large QUBO problems can be computationally challenging, especially when the matrix $Q$ is dense and the problem size grows~\cite{Nannicini_2019}. This motivates the exploration of hybrid quantum-classical algorithms, such as the one proposed in this work, as a potential means to tackle these problems more efficiently. A generic QUBO problem is both NP-hard and NP-hard to approximate, meaning that no classical, quantum, or hybrid algorithm is expected to solve these problems using polynomial resources. Instead, the promise of quantum approaches lies in the potential for more efficient approximated solutions, where the expectation is that quantum methods could offer significant computational speedups over classical algorithms to get a good approximated solution.

QUBO problems are well-suited for quantum computing because they can be naturally mapped to spin-glass Hamiltonians acting on $N$ qubits, expressed as polynomials of Pauli-$Z$ operators~\cite{hadfield2021representation}
\begin{equation}\label{eq: IsingHamiltonian}
    H = \sum_{i \in V} h_i \cdot \sigma^z_i + \sum_{(i,j)\in E} J_{ij} \cdot \sigma^z_{i} \sigma^z_{j},
\end{equation}
where the coefficients $h_i$ and $J_{ij}$ are derived from the original matrix $Q$ of the QUBO problem. This mapping is achieved by replacing the binary variables $x_i$ with the operator $\left(I-\sigma^z_i\right)/2$ where $I$ is the identity operator, and $\sigma^z_i$ denotes the Pauli-$Z$ operator acting on the $i$-th qubit. As a result, the two possible values of the binary variables $x_i$ are associated with the expectation values of this operator on the two single-qubit computational basis states. Notably, the Hamiltonian in Eq.~\eqref{eq: IsingHamiltonian} consists solely of Pauli-$Z$ operators, is diagonal in the computational basis, and has eigenstates which correspond directly to classical bitstrings. In this study, we focus on random QUBO instances by sampling the coefficients $h_i$ and $J_{ij}$ uniformly from the interval $[-1, 1]$, with four-digit precision. 

\section{ITE inspired approach}\label{sec:ITEMC}

In this section, we introduce a hybrid quantum-classical algorithm inspired by ITE to approximate the ground state of the QUBO Hamiltonian. The key idea is to construct a quantum circuit that mimics the effect of ITE using optimized unitary operations, and iteratively update the initial state based on the measurement results of the previous iteration.

ITE drives a quantum state toward the ground state by exponentially suppressing higher-energy contributions. Starting from a generic initial state $| \psi_0 \rangle$ with a non-vanishing overlap with the ground state, its imaginary time evolution converges to the ground state in the large time limit:
\begin{align}\label{eq: ITE limit}
    \ket{E_0} = \lim_{\tau\to\infty}\frac{\exp(-\tau H)\ket{\psi_0}}{\sqrt{\langle \psi_0 | {\exp(-2\tau H)} | \psi_0 \rangle}},
\end{align}
Since all terms in the Hamiltonian (Eq.~\eqref{eq: IsingHamiltonian}) commute with each other, the exact imaginary time evolution operator can be decomposed as:
\begin{equation}
    e^{-\tau H} = \prod_{(i,j)\in E} e^{-\tau J_{ij}  \sigma_i^z \sigma_j^z} \times \prod_{i\in V} e^{-\tau  h_i  \sigma_i^z}
    \label{eq:ITE_QUBO}
\end{equation}
Each exponential term in this product acts to suppress specific energy contributions, guiding the state toward a low-energy configuration.
 
However, the operators in Eq.~\eqref{eq:ITE_QUBO} are non-unitary and cannot be directly implemented on a quantum computer. Our approach is to approximate each non-unitary operator with a parameterized unitary gate that acts similarly on the current quantum state. The parameters of these gates are optimized classically to best match the imaginary-time evolution, which will be explained in Sec.~\ref {subsec: ITE_ansatz}. Although the Hamiltonian terms commute and the order of their application in the exact ITE process does not affect the final result, we will see later that operator ordering in the quantum algorithm approximation affects the performance significantly. This motivates an adaptive ordering strategy described in Sec.~\ref{subsec: gate_order}.

\begin{figure*}
\centering
\includegraphics[width=1.05\linewidth]{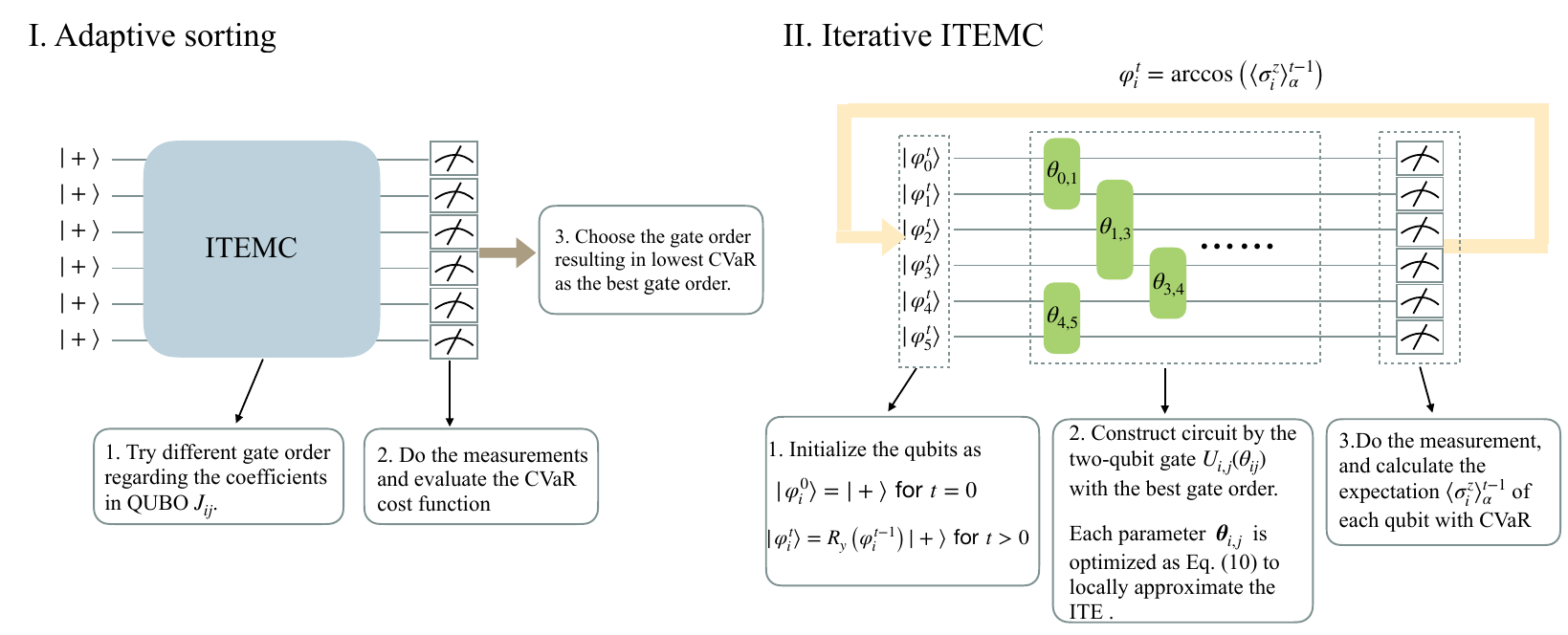}
\caption{Schematic illustration of the ITEMC algorithm.
(I) Adaptive sorting step: Different gate orderings based on the QUBO coupling coefficients $J_{ij}$ are tested, and the ordering resulting in the lowest CVaR is selected.
(II) Iterative ITEMC procedure: The qubits are initialized based on the measurement outcomes from the previous iteration. The circuit is constructed using optimized local unitaries $U_{ij}(\boldsymbol{\theta_{ij}})$ to mimic imaginary time evolution. After each iteration, the expectation values $\langle \sigma^z_i \rangle_{\alpha}^{t-1}$ are evaluated over the lowest-energy $\alpha$-fraction of sampled bitstrings, and the procedure is repeated until convergence.}
\label{fig: ITEMC_steps}
\end{figure*}

A schematic overview of the full algorithm is shown in Fig.~\ref{fig: ITEMC_steps}, which consists of two main components:
(I) an adaptive sorting step that tests different gate orderings and selects the one yielding the lowest energy as detailed in Sec.~\ref{subsec: gate_order};
(II) an iterative ITEMC that refines the state using feedback from low-energy measurement outcomes as introduced in Sec.~\ref{subsec: iterative_ITE}.

\subsection{ITE ansatz building blocks}\label{subsec: ITE_ansatz}
The ITE operator in Eq.~\eqref{eq:ITE_QUBO} consists of single-qubit and two-qubit operators, corresponding to the linear and quadratic terms of the Ising Hamiltonian. We will use two different types of unitaries for the two cases. Considering the $i$-th qubit, the single-qubit ITE operator $e^{-\tau h_i \sigma_i^z}$ acts on a general single-qubit state $\ket{\psi_{0, i}} = \alpha_i \ket{0} +  \beta_i \ket{1}$, with of $|\alpha_i|^2 + |\beta_i|^2 = 1$, as
\begin{align}
  \ket{\tilde{\psi}_{1,i}} =   e^{-\tau  h_i  \sigma_i^z}\ket{\psi_{0,i}} = e^{-\tau  h_i} \alpha_i \ket{0} + e^{\tau h_i} \beta_i \ket{1}.
\end{align}
The subindex $0$ indicates it's the initial state of this procedure, and the subindex $i$ refers to the qubit index. This non-unitary transformation can be realized, up to normalization, by applying the single-qubit rotation gate:
\begin{align}\label{eq: Yrot}
\ket{\psi_{1,i}} &= R_y(\bar{\theta}_i) \ket{\psi_{0, i}}=\exp(-i\bar{\theta}_i\sigma^y_i/2) \ket{\psi_{0, i}} 
\end{align}
with the parameter $\bar{\theta}_i =2\arctan\left(-\exp(-2\tau h_i)\right) + \pi/2$. This holds exactly for all the single-qubit terms if the initial state is an equal superposition product state $\ket{+}^{\otimes N}$. Once the corresponding rotation is applied for all qubits, the resulting state, which serves as the input to the next stage involving two-qubit operators, is:
\begin{equation}\label{eq: psi1}
    \ket{\psi_1} = \prod_{i=0}^{N-1} \otimes \ket{\psi_{0, i}}.
\end{equation}

After applying the single-qubit rotations, we approximate the effect of the two-qubit ITE operators. Specifically, the state after applying a single non-unitary operator
\begin{align}
    \ket{\tilde{\psi}_{k+1}} = e^{-\tau \cdot J_{ij} \cdot \sigma_i^z \sigma_j^z}\ket{\psi_k}, \ k \geq 1,
    \label{eq:ITE_two_qubit_terms}
\end{align}
is mimicked by the parametrized two-qubit unitary operator \cite{Chai:2024sca}\footnote{A similar ansatz can also be inspired by other physical principles, such as optimal state transfer~\cite{optstatrans_Banks2024} or counterdiabatic driving~\cite{STADAQ_Hegade2021, BFDCQO_Cadavid2025}.}
\begin{equation}\label{eq: SIA-YZ}
  \ket{\psi_{k+1}} =   U_{ij}(\boldsymbol{\theta}_{ij})\ket{\psi_k} = e^{-i (\theta_{ij, 1} \cdot \sigma_i^z \sigma_j^y + \theta_{ij, 0} \cdot \sigma_i^y \sigma_j^z)/2}\ket{\psi_k},
\end{equation}
where the parameters $\boldsymbol{\theta}_{ij} = \{ \theta_{ij, 0}, \theta_{ij, 1}\}$ are optimized to maximize the overlap between the ideal ITE-evolved state and that obtained by quantum circuit:
\begin{equation}\label{eq: f-local-op}
    \begin{aligned}
        &\max_{\boldsymbol{\theta}_{ij}}f_{\tau, k+1}(\boldsymbol{\theta}_{ij})= \max_{\boldsymbol{\theta}_{ij}} \bra{\tilde{\psi}_{k+1}} \ket{\psi_{k+1}}\\
        &=\max_{\boldsymbol{\theta}_{ij}}  \bra{\psi_{k}} e^{-\tau \cdot J_{ij} \cdot \sigma_i^z \sigma_j^z} \cdot e^{-i (\theta_{ij,1} \cdot \sigma_i^z \sigma_j^y + \theta_{ij,0} \cdot \sigma_i^y \sigma_j^z)/2} \ket{\psi_{k}}
    \end{aligned}
\end{equation}
 This procedure is carried out sequentially for all two-qubit operations in Eq.~\eqref{eq:ITE_QUBO}, using the current quantum state \( \ket{\psi_k} \) obtained after applying all previously optimized gates. Once all unitaries are fixed, the full ITE-inspired ansatz is defined as:
\begin{equation}\label{eq: SIA-ansatz}
    \ket{\psi^0(\boldsymbol{\bar{\theta}})} :=
    \prod_{(i,j)\in E} U_{ij} (\boldsymbol{\bar{\theta}}_{ij}) \times \prod_{i \in V} R_y(\bar{\theta_i}) \ket{+}^{\otimes N},
\end{equation} 
where $\bar{\theta}_i$ and $\bar{\boldsymbol{\theta}}_{ij}$ denote the optimized parameters and this state serves as the output of a single iteration of the ITEMC. If optimization of the parameters $\boldsymbol{\theta}$ leads to a good approximation of the ITE, the energy expectation value in the state $\ket{\psi^0(\boldsymbol{\theta})}$ will be significantly lower than in the initial state. 

Importantly, in the optimization procedure, the cost functions $f_{\tau,k+1}$ can be estimated from the expectation value of a few one-qubit and two-qubits observables, which only need be measured once and don't need repeated during the optimization (see Appendix \ref{Appendix A} for details). Specifically, once these expectation values are estimated over a given state $\ket{\psi_k}$, the corresponding parameter optimization in Eq.~\eqref{eq: f-local-op} becomes a purely classical problems.  With precision $\epsilon$, the number of shots for evaluating these expectations is $\mathcal{O}(1/\epsilon^2)$. The state $\ket{\psi_k}$ needs to be prepared iteratively for each operator in Eq.~\eqref{eq:ITE_QUBO} to estimate all parameters in Eq.~\eqref{eq: SIA-ansatz}. The number of operators is determined by the Hamiltonian, which has $M = N + \frac{N(N-1)}{2}*d$ terms for a graph density $d$. The computational cost for the optimization of the entire circuit is therefore at worst $\mathcal{O}(M/\epsilon^2) \leq \mathcal{O}(N^2/\epsilon^2)$. We name this approach \emph{ITEMC by measuring}. Ref.~\cite{Chai:2024sca} also proposed a more simplified way to approximate Eq.~\eqref{eq: f-local-op}, substituting all the states $\ket{\psi_k}$ with the product state $\ket{\psi_1}$ in Eq.~\eqref{eq: psi1}, which can provide an approximation of Eq.~\eqref{eq: f-local-op} with error $\mathcal{O}(\tau)$. In this way, the necessary Pauli expectations in Eq.~\eqref{eq: f-local-op} can be calculated analytically in the product state, and there is no additional cost to determine the optimal parameters $\bar{\boldsymbol{\theta}}$ in the circuit. We name it as \emph{ITEMC by approximation}.

After the construction of the whole circuit with the optimal parameters, we can measure the state $\ket{\psi^0(\boldsymbol{\theta})}$ to estimate the expectation of the energy, if needed, at a given precision $\epsilon$ with $\mathcal{O}(M/\epsilon^2) \leq \mathcal{O}(N^2/\epsilon^2)$ shots. It is important to note here that the estimation of the energy is not part of the optimization procedure. This makes our method computationally cheaper than variational methods such as VQE. In VQE, the energy of the entire Hamiltonian must be evaluated at every optimization step, leading to a cost that scales as $\mathcal{O}(s M^2 / \epsilon^2) \leq \mathcal{O}(s N^4 / \epsilon^2)$, where $s$ is the number of iterations and $\mathcal{O}(M)$ evaluations are required per iteration to calculate gradients for all parameters~\cite{Peruzzo:2013bzg}. A comparison of the required measurements between our algorithm and VQE are shown in Table.~\ref{tab: resource}.
\begin{table}[h]
    \centering
    \begin{tabular}{|c|c|c|c|}
    \hline
                                   & $\bar{\boldsymbol{\theta}}$ & CVaR & total \\
    \hline
         \emph{ITEMC by measuring} &  $\mathcal{O}(M/\epsilon^2)$ & $\mathcal{O}(M/\epsilon^2)$ & $s^{\prime} * \mathcal{O}\left( 2M/\epsilon^2 \right)$ \\
    \hline
         \emph{ITEMC by approximation} &  0 & $\mathcal{O}(M/\epsilon^2)$ & $s^{\prime} * \mathcal{O}\left( M/\epsilon^2 \right)$\\
    \hline
         VQE                        & $\backslash$ & $\mathcal{O}(M/\epsilon^2)$ & $s * \mathcal{O}\left( M^2/\epsilon^2 \right)$ \\
    \hline
    \end{tabular}
    \caption{Resource estimates for number of measurements. The column labeled $\bar{\boldsymbol{\theta}}$ refers to the cost of determining circuit parameters via measurements of local Pauli operators. In the \emph{ITEMC by approximation} variant, these parameters are estimated analytically without measurements, while in VQE they are not explicitly required. The CVaR column indicates the measurement cost for evaluating the expectation value of the Hamiltonian to precision $\epsilon$. The final column shows the total resource estimate, where $s^{\prime}$ denotes the number of iterations in ITEMC, which is typically much smaller than the number of iterations $s$ required by VQE, as demonstrated in our numerical results.}
    \label{tab: resource}
\end{table}

\subsection{Iterative ITE}\label{subsec: iterative_ITE}
The imaginary time evolution in Eq.~\eqref{eq: ITE limit} converges to the ground state as $\tau \to \infty$, but this process increases the correlations between qubits. As shown in Ref.~\cite{Motta_2019}, to accurately approximate the evolution operator in Eq.~\eqref{eq:ITE_QUBO} with a quantum circuit,  unitary gates acting on increasingly nonlocal subsets of qubits are required as correlations grow, leading to deeper circuit depths. A better strategy is instead to consider smaller values of $\tau$, update the initial state and repeat the ITE procedure until convergence.

We begin with the uniform superposition state $\ket{+}^{\otimes N}$, where the expectation values of each Pauli-Z operator is initially zero:
\begin{align}
   \bra{+}^{\otimes N} \sigma^z_i \ket{+}^{\otimes N} = 0.
\end{align}
After one ITE-mimicking circuit, the resulting state $\ket{\psi^0(\boldsymbol{\theta})}$ in Eq.~\eqref{eq: SIA-ansatz} have nonzero $\sigma_i^z$ expectation values:
\begin{align}\label{eq: expz}
   \bra{\psi^0} \sigma^z_i  \ket{\psi^0} = \cos(\varphi_i^0),
\end{align}
where the superscript represents the iteration index. Thus, the parameters $\varphi_i^0$ quantify how far each qubit must deviate from the equal superposition state to approach the ground state. This information can be used as initial condition for a subsequent iteration of the ITE mimicking procedure: for the first iteration step $t=0$, the quantum state is initialized in an equal superposition by applying Hadamard gates to each qubit, while for subsequent iteration steps $t$, the initialization is achieved by applying $R_y(\varphi_i^{t-1})$  rotations. The angles $\varphi_i^{t-1}$ are given by the $\sigma^z_i$ expectation values in Eq.~\eqref{eq: expz} obtained from the previous iteration. This feedback loop guarantees that the energy expectation value will decrease significantly after a few iterations. \\

In the following, both the energy and the $\sigma^z_i$ expectation values are computed with the conditional value at risk (CVaR)~\cite{ACERBI20021487} 
to focus on low-energy states sampled during measurements. 
The CVaR is defined as the average of the lowest $\alpha$-fraction of the energy eigenvalues:
\begin{equation}
    \text{CVaR}_{\alpha} = \frac{1}{\lceil \alpha S \rceil} \sum_{k = 1}^{\lceil \alpha S \rceil} E_k.
    \label{eq:CVaR}
\end{equation}
with eigenvalues $\{E_1 \leq E_2 \leq \cdots \leq E_S\}$ obtained from $S$ measurements sorted in ascending order. Additionally, the corresponding best $\alpha S$ bit strings are used to estimate the expectation of Pauli Z operator, which we denote as $\langle \sigma^z_i \rangle_{\alpha}$.
CVaR reduces to the standard mean energy estimate for $\alpha =1$.
Throughout this work, we use $\alpha = 0.01$, unless otherwise specified. 

Before showing our results, we emphasize again that our method is significantly cheaper than other variational methods such as VQE or QAOA, as the optimization procedure involves the computation of single-qubit and two-qubits expectation values only, which can be evaluated with much fewer shots than the full energy needed in the other methods. This estimate holds as long as the number of iterations of the ITEMC does not increase significantly with the system size, which we will show numerically below to be indeed the case.

\subsection{Circuit implementation}\label{subsec: gate_order}
In the circuit implementation of the algorithm above, the circuit begins with single-qubit rotation gates, followed by all possible 2-qubit gates as in Eq.~\eqref{eq: SIA-YZ}.  In the implementation of the 2-qubit gates, one possibility is to design a circuit with minimal depth, without considering the effect of gate ordering. The exact circuit architecture will, of course, depend on the connectivity of the specific hardware. An example of a brickwork structure with linear connectivity can be found in Appendix B of \cite{Chai:2024sca}.\\
 As the coupling coefficients $J_{ij}$ associated with each gate can vary widely, gate ording can affect the performance. In order to take this into account, for a given instance, we consider four sorting strategies for the coupling coefficients $J_{ij}$, sorting by increasing or decreasing value, and by increasing or decreasing absolute value. The order of the two-qubit gates in the circuit implementation follows the chosen sorting scheme. As a consequence, these circuits generally exhibit a greater depth, scaling as $\mathcal{O}(N^2)$ for $N$ qubits, but often yield better performance. To leverage this, we adopt an adaptive sorting strategy. In the first iteration of ITEMC, we run the circuit once for each of the five ordering options (including the original unsorted order), and select the one that yields the lowest energy. All subsequent ITEMC iterations are then performed using this best-performing ordering. This adaptive approach increases the quantum measurement cost slightly. Each iteration step requires $\mathcal{O}( M/\epsilon^2)$ shots, and evaluating all five sorting options in the first step adds an additional cost of $\mathcal{O}(5 \times M/\epsilon^2)$ shots.\\

\section{Numerical simulations results}\label{sec:results}

We evaluate the performance of the ITEMC through numerical simulations on random QUBO instances with qubit numbers ranging from 10 to 150.  We use the approximation ratio as the performance metric, which is defined as $\mathrm{CVaR}_{\alpha} / E_{\mathrm{opt}}$, with $\mathrm{CVaR}_{\alpha}$ the optimized cost function obtained by ITEMC and $E_{\mathrm{opt}}$ the ground state energy. $E_{\mathrm{opt}}$ is computed exactly by brute force for up to 22 qubits; for larger sizes, it's estimated by simulated annealing.

For small problem sizes up to 22 qubits, various graph densities are considered, including 3-regular graphs, intermediate densities (0.5, 0.7, 0.8, 0.9, 0.95), and complete graphs. For each combination of problem size and graph density, 400 random instances are generated. We perform statevector simulations with both infinite and finite shots. In the latter, 1,000 measurements are used to evaluate the Pauli expectations to get the optimal parameters in Eq.~\eqref{eq: f-local-op}, and 10,000 measurements are used to estimate the final $\mathrm{CVaR}_{\alpha}$ and $\langle \sigma_i^z\rangle_{\alpha}$. It is important to emphasize that the more costly energy evaluation is performed only once, after the circuit parameters have been optimized. The classical optimization to maximize the Eq.~\eqref{eq: f-local-op} is carried out using Sequential Least Squares Programming (SLSQP) optimization, with a maximum of 10000 iterations.  Importantly, the Pauli expectations in Eqs.~\eqref{eq: f-detail} serve only as fixed coefficients in the cost function and are computed once per ITEMC iteration, they do not need to be remeasured during classical optimization by SLSQP.

For larger problem sizes (60 to 150 qubits), we simulate ITEMC using the matrix product state (MPS) backend in Qiskit. In this case, we restrict to 3-regular graphs, as simulating denser graphs becomes computationally expensive due to rapid entanglement growth, as we will show later.

The hardware runs are demonstrated on IBM’s quantum devices for 40, 60, and 80 qubits,

\subsection{Statevector simulation on small systems}
\begin{figure}[htp!]
    \centering
    \includegraphics[width = 0.4\textwidth]{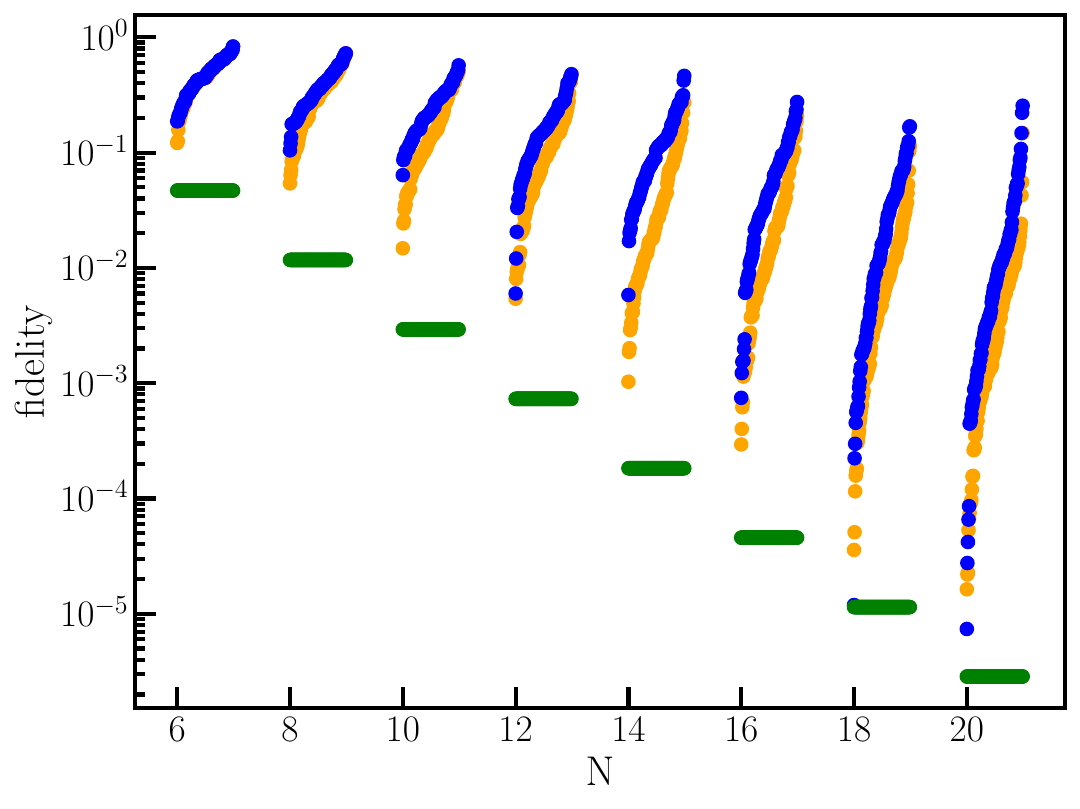}
    \caption{Log-scale plot for the probability of the best three solutions averaged among 100 instances from $N=6$ to $N=20$ qubits at $\tau = 0.6$. The blue and orange points correspond to the algorithm with and without adaptive-sorting, respectively. The fidelity is computed with respect to the 3 lowest energy solutions. The green line represents the fidelity of the uniform superposition which is $3/2^N$.}
    \label{fig: cvarsort_best3_tau06}
\end{figure}

In this subsection, we will show our numeric results on small problem sizes, which can be simulated efficiently and allow for a comprehensive exploration of algorithmic performance. Firstly, we benchmark the impact of gate ordering in the circuit. It is worth noting that no single sorting method consistently outperforms the unsorted circuit across all instances. The optimal sorting arrangement is instance-dependent, which motivates the use of the adaptive sorting strategy: we test all sorting schemes in the first iteration and select the one that yields the lowest energy for subsequent iterations. Figure~\ref{fig: cvarsort_best3_tau06} shows the probability of the best three solutions in the state prepared in the first iteration $\ket{\psi^0(\boldsymbol{\bar{\theta}})}$ (Eq.~\eqref{eq: SIA-ansatz}).  For each problem size, the results are averaged over 100 random instances of complete graphs at $\tau=0.6$. We compare the performance of ITEMC with unsorted gate ordering (organge) and with adaptive sorting (blue), against a baseline given by the uniform superposition state (green line). Both ITEMC methods show a significant improvement over random sampling, with the adaptive sorting strategy yielding better results by reducing the number of poorly performing instances.

\begin{figure}[htp!]
    \centering
        \includegraphics[width = 0.75\linewidth]{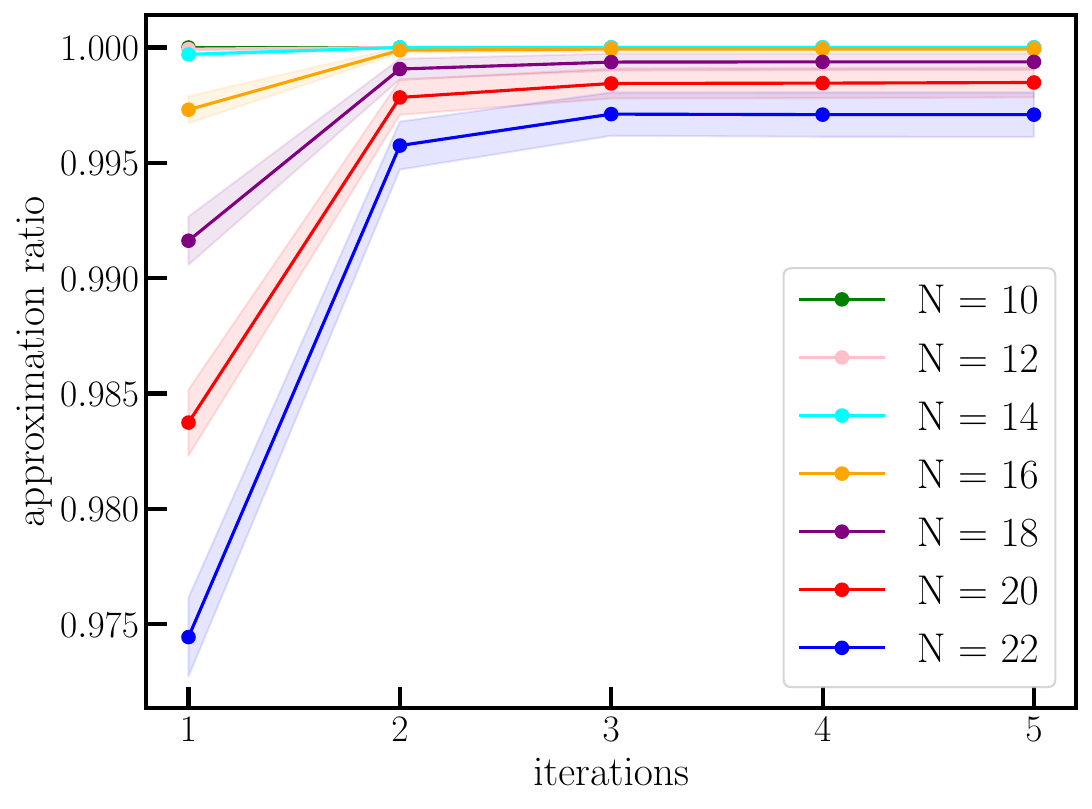}
     \caption{Approximation ratio as a function of iteration number for complete graph QUBO instances using ITEMC with adaptive sorting. Simulations are ideal (noise-free) and span various system sizes, the results of each problem size are averaged among 400 random instances. The computation is performed with $1000$ shots for the Pauli expectation during the optimization and $10000$ shots for the evaluation of the CVaR energy. Parameters are set to $\alpha = 0.01, \tau = 0.3 $, and error bars represent 95\% confidence intervals. The approximation ratio saturates after a few iterations, motivating our choice to fix the maximum number of iterations to 5 in subsequent analyses. }
    \label{fig: ar_iter_N}
\end{figure}

\begin{figure}[htp!]
    \centering
    \includegraphics[width = 0.8\linewidth]{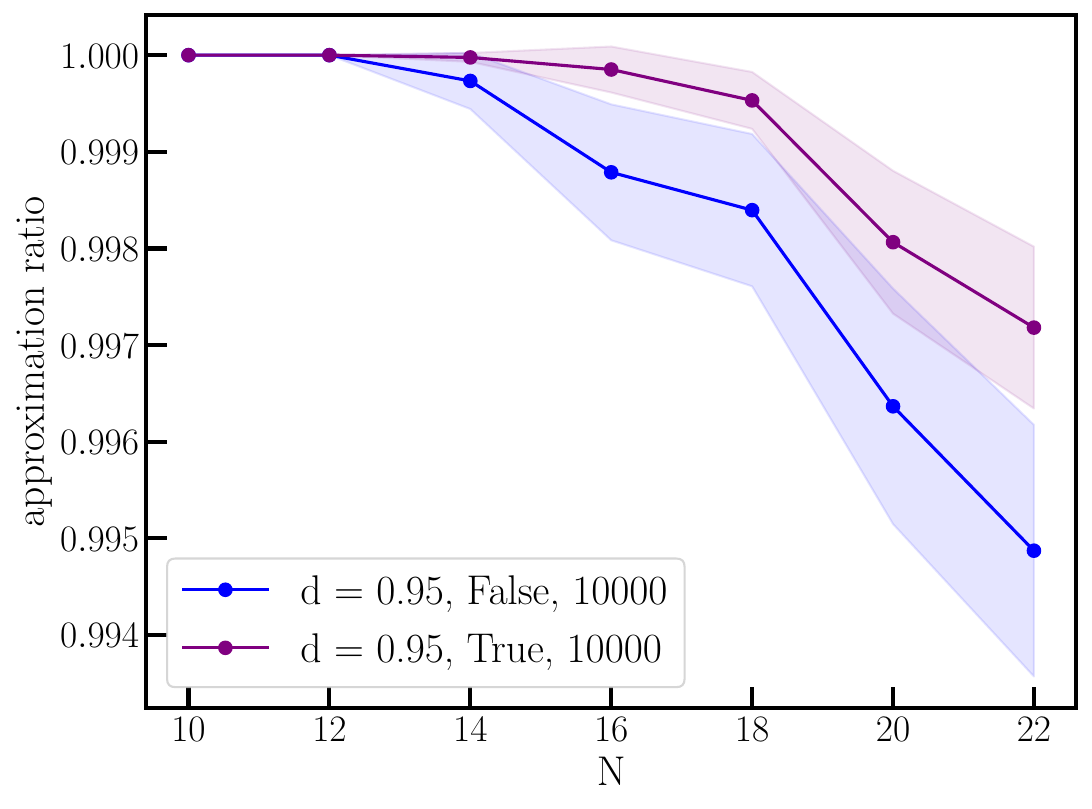}
    \caption{The approximation ratio with 95\% confidence level of the ITEMC solution for 10 to 22 qubits and 0.95 density graphs. The results with adaptive sorting are labeled with ``True" in the legend, and random sorting are labeled with ``False". The results of each problem size are averaged among 400 random instances. The shot simulation is carried out with 1000 shots for the optimization procedure and 10000 for the evaluation of the final state only.}
    \label{fig: ar_truevsfalse_N}
\end{figure}

\begin{figure}[htp!]
    \centering
    \includegraphics[width = 0.8\linewidth]{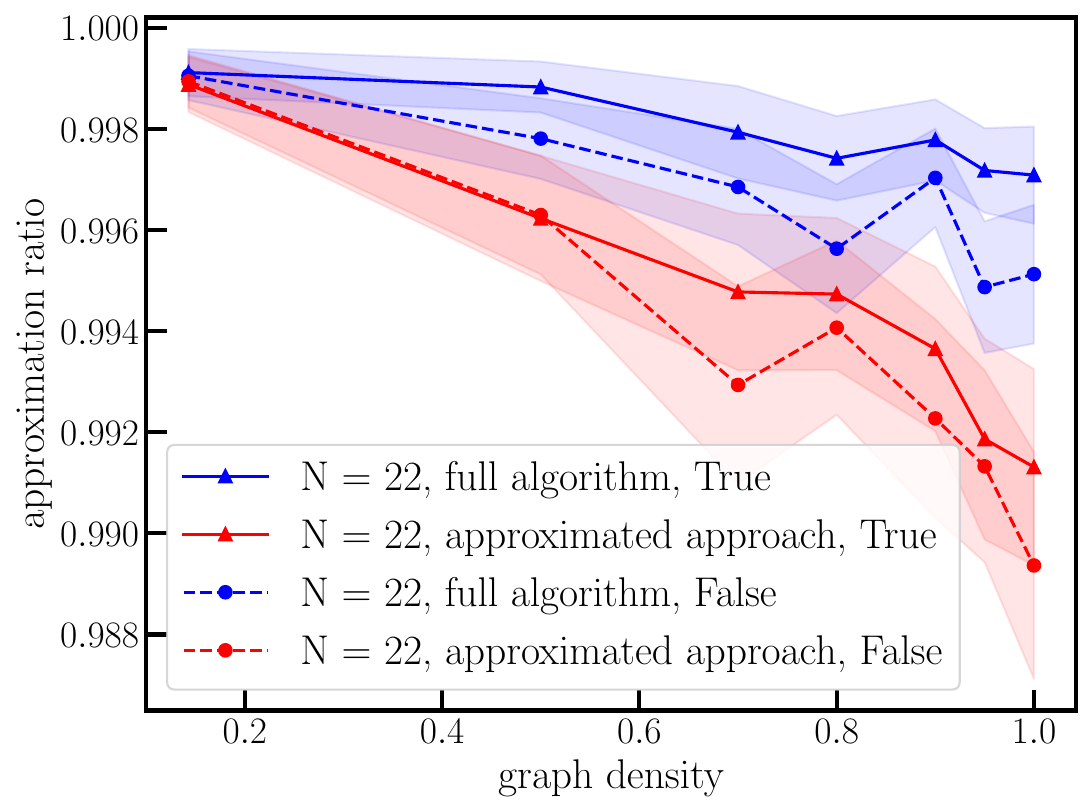}
    \caption{Approximation ratio with 95\% confidence level for different graph densities. All results are for 22 qubits, averaged over 400 random instances per density. The lowest density corresponds to a 3-regular graph. The simulation is carried out with 1000 shots for the optimization procedure and 10000 for the final evaluation only. The approximated approach \emph{ITEMC by approximation} shows comparable performance at low density, while it fails to capture the whole dynamics at higher density, resulting in a worse approximation ratio. In addition, the solid lines and dotted lines represent the results from adaptive sorting and random sorting respectively.}
    \label{fig: full_vs_ideal_densities}
\end{figure}

After selecting the best-performing gate order, we continue with the iterative procedure. As shown in Figure~\ref{fig: ar_iter_N}, the approximation ratio increases rapidly over the ITEMC iterations typically converges within five iterations. This convergence behavior remains stable across different problem sizes. We will later show in Sec.~\ref{sec:tensornetworks} that this favorable scaling continues for larger systems using MPS simulations. Figure~\ref{fig: ar_truevsfalse_N} shows how the approximation ratio scales with system size. While the ratio slightly decreases as the number of qubits increases, it remains high, around $0.997$ for the best gate order (red curve). Performance can be further improved by using a smaller CVaR parameter $\alpha$ and more measurement shots, which will be demonstrated in the next subsection. For comparison, Fig.~\ref{fig: ar_truevsfalse_N} also shows results using random gate ordering (blue curve). Although performance is slightly lower (approx. 0.995), it still provides a good approximation. Figure~\ref{fig: full_vs_ideal_densities} highlights that while gate ordering affects results, the difference is negligible at low graph densities and becomes more relevant for high density. The magnitude of this effect also depends on the value of $\tau$ (see Appendix~\ref{appendix: tau}). The choice may depend on the problem-specific need for precision. It would be interesting to identify patterns in optimal gate orderings. A promising direction would be to use machine learning to predict effective gate sequences, which can save time by skipping adaptive sorting step. For simplicity, results shown without explicit labels will correspond to ITEMC with adaptive sorting.

We explore the algorithm's performance across various graph densities in Fig.~\ref{fig: full_vs_ideal_densities}. Consistent with prior observations~\cite{Nannicini_2019, role_entangle_2021}, our result indicates the dense graph is more challenging for optimization. Nevertheless, with adaptive gate sorting, ITEMC maintains a high approximation ratio, around $0.997$ even for complete graphs. The performance could be further improved by using a smaller CVaR as shown in the next subsection.

\begin{figure}[htp!]
    \centering
    \includegraphics[width = 0.8\linewidth]{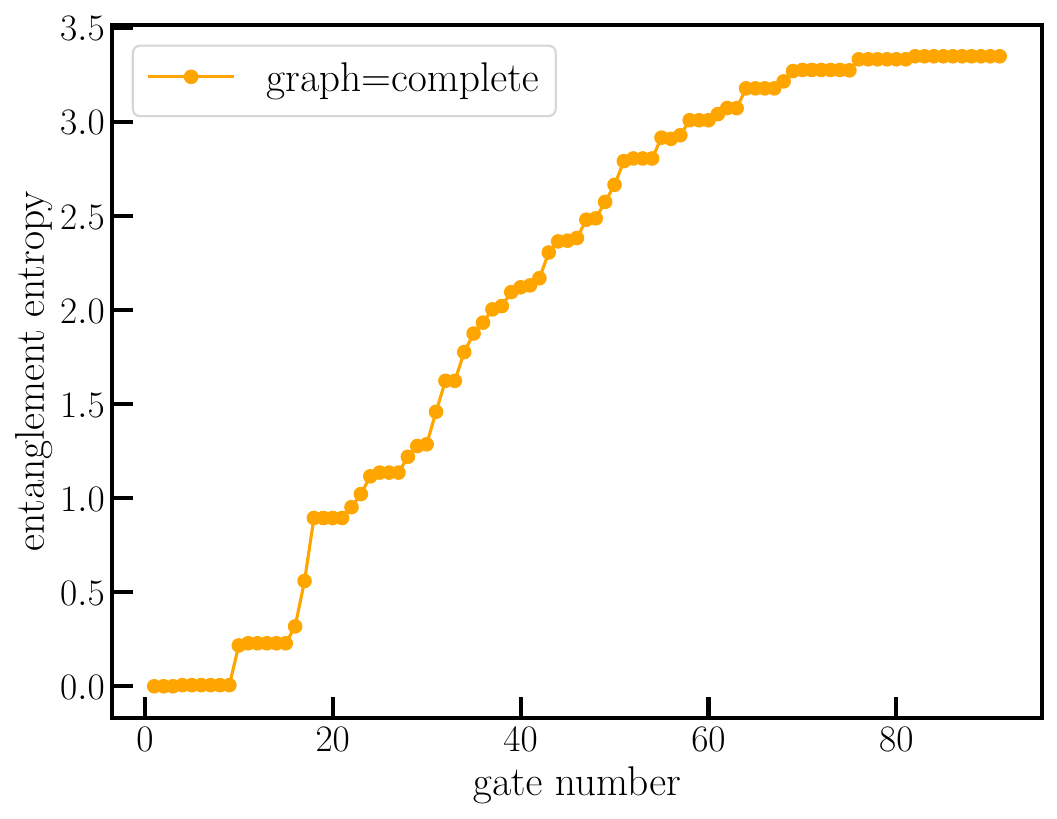}
    \caption{Evolution of the entanglement entropy along the ITEMC for a random instance with 14 qubits for a complete graph. Plateaus correspond to gates acting only on qubits that are traced out.}
    \label{fig: entropyvsgates}
\end{figure}

We also explore the performance of \emph{ITEMC by approximation}, where all parameters are estimated in the product state.  Since the entanglement in the state grows with the two-qubits gate in the circuit as shown in Fig.~\ref{fig: entropyvsgates}, the circuits corresponding to denser graphs will deviate more from the product state form. We expect this approximation approach to perform well for sparse graphs but it may degrade for dense graphs. Figure~\ref{fig: full_vs_ideal_densities} compares the performance of the full and approximated ITEMC for 22 qubits across various graph densities. As expected, the full ITEMC shows a mild decline in approximation ratio with increasing density, the drop is gradual and performance remains high in the complete graph. In contrast, the approximation approach performs similarly to the full algorithm on sparse graphs but diverges significantly as density increases. This confirms that parameter estimation under the product state assumption is an efficient and reasonable approximation for sparse graphs, allowing reduced computational cost, but becomes unreliable for dense graphs due to increasing entanglement.
\begin{figure}[htp!]
    \centering
    \includegraphics[width = 0.8\linewidth]{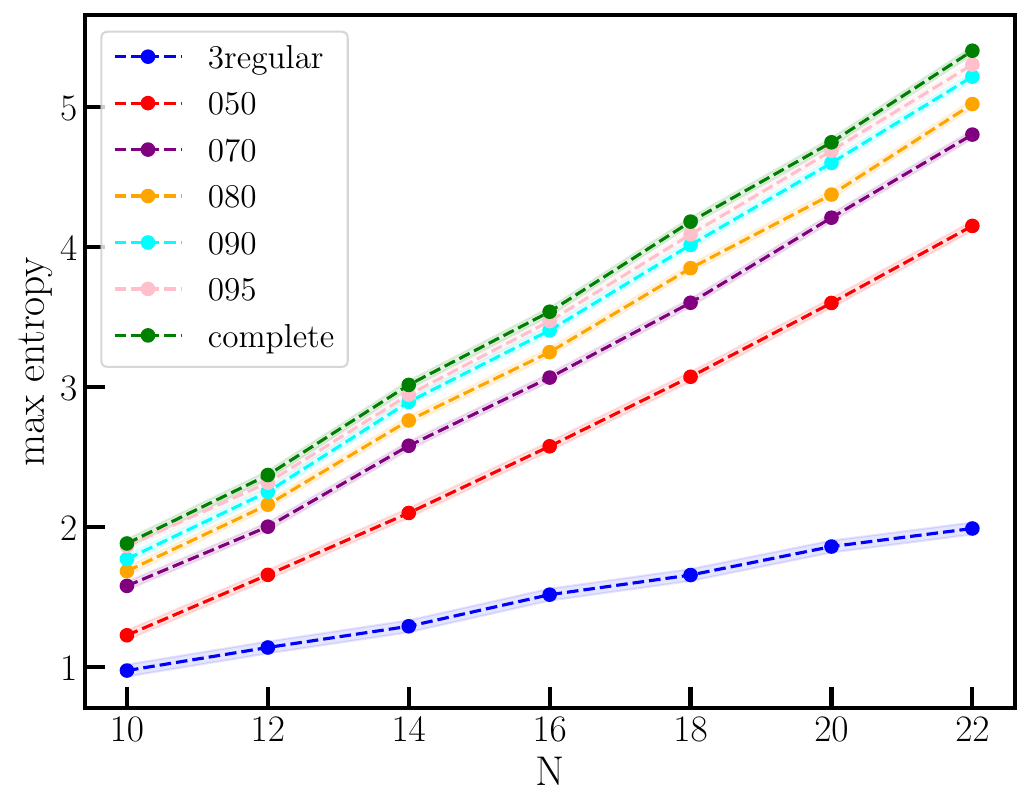}
    \caption{Maximal entropy generated by the ITEMC within 5 iterations versus the number of qubits. The scaling is linear for density 0.5 to 1, and milder for 3-regular graph.}
    \label{fig: max_entropy}
\end{figure}

Fig.~\ref{fig: max_entropy} shows the scaling of the maximum entanglement entropy generated by the ITEMC as a function of the number of qubits. As discussed above, the circuit corresponds to a denser graph will yield more entanglement. For graph densities between 0.5 and fully connected graphs, the entropy scales linearly with system size, while for 3-regular graphs, the scaling is sublinear or linear with a small slope~\cite{note_entropy_scaling}. These findings indicate that while the ITEMC achieves optimal performance on 3-regular graphs, such instances might remain classically simulable using MPS techniques. In contrast, QUBO instances with densities above 0.5 are much harder to simulate classically due to high entanglement, where the ITEMC continues to exhibit strong performance. This highlights the robustness and practical relevance of our method for dense optimization problems using quantum computing.

\subsection{MPS simulation on large system}\label{sec:tensornetworks}
\begin{figure}
    \centering
    \includegraphics[width=0.95\linewidth]{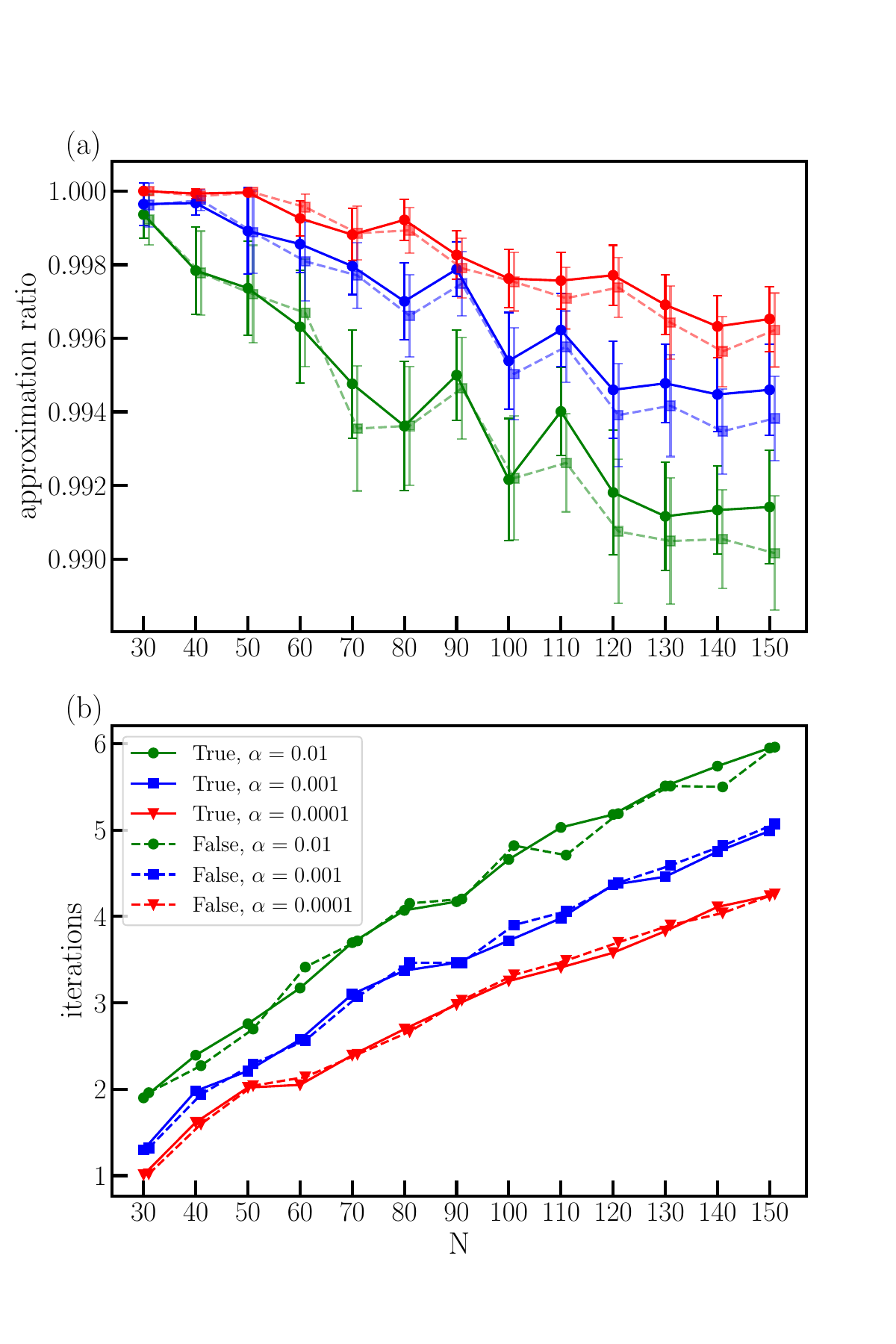}
    \caption{Scaling of approximation ratio and iterations for ITEMC on the 3-regular QUBO instances. (a) Approximation ratio at different problem sizes from 30 to 150 qubits. (b) Number of iterations required for convergence. For simplicity, we only show the results for every 20 qubits, and the result for each problem size is averaged among 100 random instances.}
    \label{fig: mps_3reg}
\end{figure}

We extend our numerical simulations to larger system sizes, ranging from $30$ to $150$ qubits, using the MPS backend of Qiskit, with a maximal bond dimension of 100. Due to the rapid growth of the entanglement entropy for denser graphs, we are only able to do this simulation for 3-regular graphs. For each problem size, we generate 100 random 3-regular graphs and define the corresponding QUBO problem according to Eq.~\eqref{eq: IsingHamiltonian}. The reference ``optimal" solutions here are obtained from simulated annealing~\cite{dwave-neal}. Fig.~\ref{fig: mps_3reg}(a) shows the results obtained from the ITEMC method by the approximated approach explained above. We tested different CVaR coefficients and the two sorting strategies, for all setups, the approximation ratio remains high across all problem sizes. Although it slightly decreases for larger problem sizes, the approximation ratio remains around $99$ to $99.6\%$ for even 150 qubits. Using adaptive sorting and smaller CVaR coefficients can improve the performance. In particular, a smaller $\alpha$ not only improves the approximation ratio significantly, but also reduces the number of iterations needed for convergence as shown in Fig.~\ref{fig: mps_3reg}(b), where convergence is defined as the relative cost function change smaller than $1e-4$. Since smaller CVaR values require more measurements to maintain sufficient statistical accuracy, we fix $\mathrm{shots} \times \alpha = 100$ to ensure reliable estimation of both the cost function and the single-qubit $\sigma^z$ expectation values. \\Fig.~\ref{fig: mps_3reg}(b) shows how the number of iterations scales with problem size. The required number of iterations increases only slowly with the system size, and problems with 150 qubits converge in only 6 steps. Since we use the approximated approach, there is no extra measurements needed to determine the parameters in the circuit, and the quantum resource cost is minimal. Specifically, only 5 circuits are required to determine the optimal gate ordering, and approximately 6 circuit executions are needed to perform the iterative optimization. Thus, the total number of circuit executions required is around 11, which is significantly lower than the number of evaluations typically needed in variational algorithms such as VQE or QAOA.

\section{Hardware run}\label{sec:hardware}
For hardware testing, the ITEMC algorithm was run on IBM’s \texttt{ibm\_fez} superconducting quantum processor, a 156-qubit device based on the Heron r2 architecture.
At the time of the experiments (April 10, 2025), the device calibration showed an average readout error of approximately $2.07*10^{-2}$, with a median readout error of $8.79*10^{-3}$.  The two-qubit gate errors exhibited a mean of about $1.34*10^{-1}$ and a median of $4.38*10^{-3}$. The device had an average $T_1$ (energy relaxation) time of 157.6 $\mu s$ and a median of 159.4 $\mu s$, and an average $T_2$ (dephasing) time of 110.4 $\mu s$ with a median of 113.0 $\mu s$ across all qubits.

The tests are run for a single QUBO instance defined on a 3-regular graph for 40, 60 and 80 qubits. 
The CZ circuit depth is approximately 45, 55, or 65 for the three cases, and varies by a few units at each iteration due to changes in the qubit layout, which are made by an AI-powered transpiler to minimize experimental error at the time of execution. To suppress experimental errors, we also apply Pauli twirling to measurement instructions~\cite{PauliTwirling_Wallman2016} and use XY4-type dynamical decoupling to suppress decoherence during idle periods.\\

As for the MPS simulations for large system size, here we also use the \emph{ITEMC by approximation}. Thus, the approximated optimal parameters are estimated analytically, and the quantum hardware is used only to implement the ITEMC for each iteration and measure the $\sigma^z$ expectation values and the CVaR energy of the resulting quantum state. The quantum circuit is initially run on the hardware 5 times, one per sorting scheme respectively, to select the best sorting, corresponding to the lowest energy. Then, the quantum circuit with the selected best sorting is run iteratively on the hardware for 10 iterations, each time computing the $\sigma^z$ expectation values, used for the initialization of the subsequent iteration. \\

\begin{figure}[htp!]
    \centering
    \includegraphics[width = 0.8\linewidth]{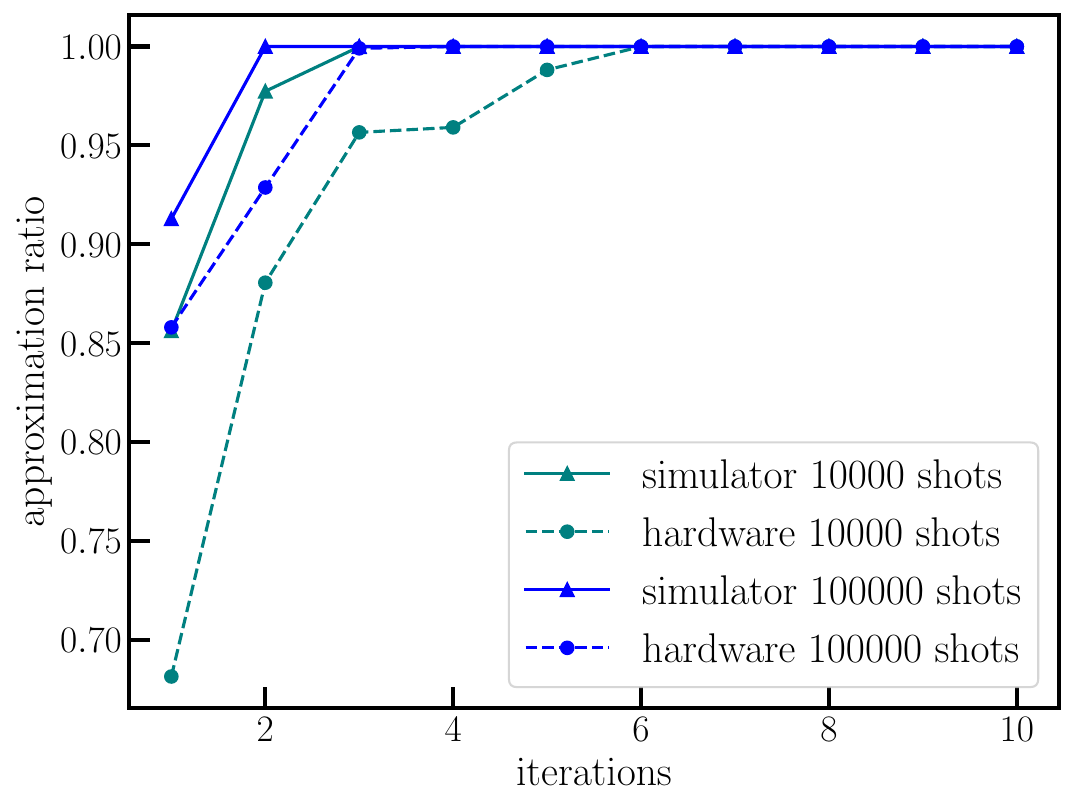}
    \caption{Approximation ratio versus the number of iterations of a single 3-regular graph instance for 40 qubits. The simulation and the hardware implementation are run with MPS backend and \texttt{ibm\_fez} respectively. In the hardware run, convergence is reached after 7 iterations with 10000 shots and 4 iterations with 100000 shots, respectively.}
    \label{fig: hardware_40}
\end{figure}

\begin{figure}[htp!]
    \centering
    \includegraphics[width = 0.8\linewidth]{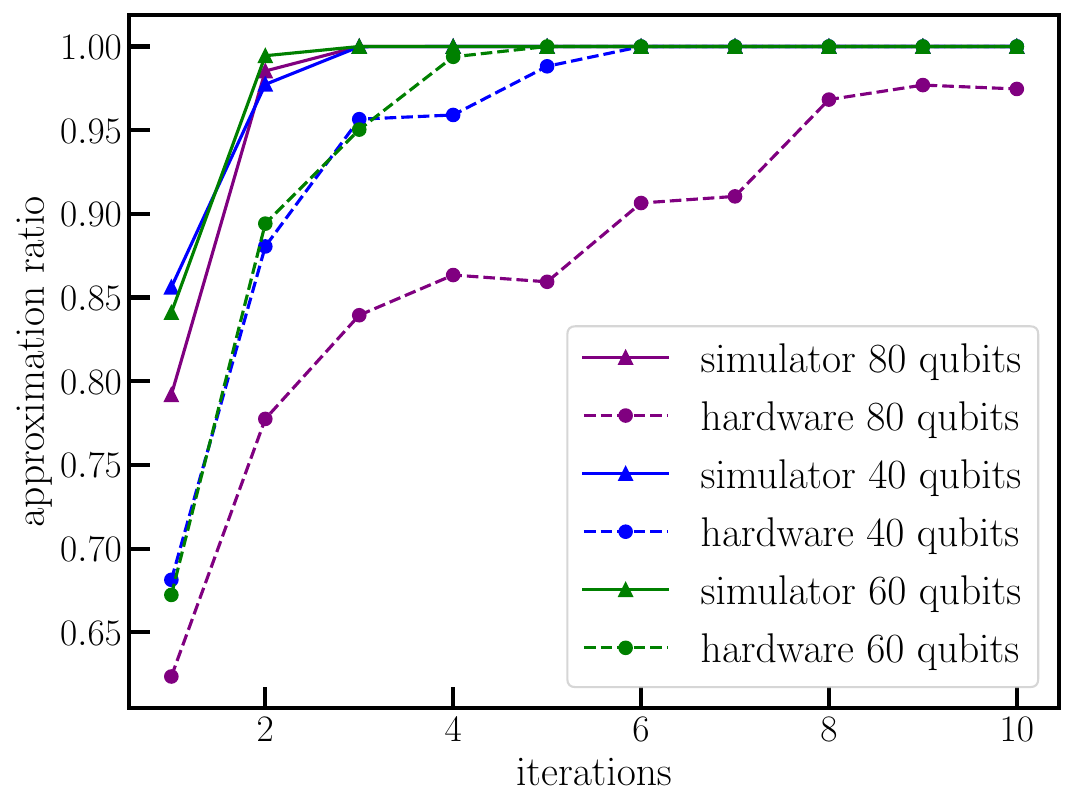}
    \caption{Approximation ratio versus the number of iterations of a single 3-regular graph instance for 40, 60 and 80 qubits. The simulation and the hardware implementation are run with MPS backend and \texttt{ibm\_fez}. The number of shots per iteration is 100000.}
    \label{fig: hardware_100000}
\end{figure}

Fig.~\ref{fig: hardware_40} shows the results of the hardware run compared to the MPS simulation for 40 qubits. The number of iterations is fixed at 10, but convergence is reached after 7 iterations for 10,000 shots and after 4 iterations for 100,000 shots. Note that in the former case, even though more iterations are needed, the total number of shots is lower.
In Fig.~\ref{fig: hardware_100000}, the number of shots is fixed at 100,000, and the hardware test is extended to instances with 60 and 80 qubits.
The approximation ratio and the fidelity reached in the hardware runs are resumed in Table \ref{tab: hardware}. We highlight here that, while the fidelity of the solution obtained for 80 qubits is rather low (being $2*10^{-5}$), the number of shots is $10^{5}$, which implies that the ITEMC was able to find the best solution at least twice during the computation. The same fidelity is also obtained for the second and third lowest solutions in this case. For 40 and 60 qubits, the ground state is obtained with high fidelity, with the second-highest fidelity being more than an order of magnitude smaller.

\begin{table}[]
    \centering
    \begin{tabular}{|c|c|c|c|c|c|}
    \hline
                              qubits    & ar & fidelity & shots per iteration & iterations & tot shots \\
    \hline
         40  &  1 & 0.759 &$10^4$ & 7 & $1.2*10^5$ \\
    \hline
         40  &  1 & 0.763 &$10^5$& 4 & $9*10^5$ \\
    \hline
         60  &  1 & 0.594 &$10^5$& 5 & $10^6$\\
    \hline
         80                       & 0.974  & $2*10^{-5}$ &$10^5$& 9 & $1.4*10^6$ \\
    \hline
    \end{tabular}
    \caption{Hardware run results. The fidelity is computed with respect to the best solution found with simulated annealing. The approximation ratio (ar) is computed with CVaR with $\alpha = 0.01$ and $\alpha = 0.001$ for $10^4$ and $10^5$ shots respectively. The computation of the total number of shots includes the 5 iterations needed for selecting the best sorting plus the iterations needed for reaching convergence.}
    \label{tab: hardware}
\end{table}

\section{Summary and outlook}\label{sec:summary}
This paper proposes a hybrid quantum-classical algorithm for solving QUBO problems using an ITEMC. Inspired by imaginary time evolution, which naturally drives quantum states toward ground states, the method constructs a quantum circuit composed of parameterized single- and two-qubit gates that mimic this evolution. Unlike standard variational approaches such as VQE or QAOA, ITEMC enables efficient parameter determination either through lightweight classical optimization or by approximated analytical calculations without requiring additional quantum measurements. An important innovation in our method is the incorporation of a preliminary sorting step, where we optimize the arrangement of quantum gates by selecting the best sorting scheme from five possible configurations. This adaptive sorting process further enhances the algorithm's performance without introducing significant overhead. Unlike variational methods that require full Hamiltonian expectation measurements at each step, ITEMC only uses local (one- and two-qubit) expectation values. These can be measured once per iteration, leading to a much lower shot complexity. The algorithm uses CVaR as the cost function to focus on low-energy configurations and introduces an iterative scheme where the initial state of each iteration is informed by measurements from the previous step.

Classical simulations are performed across 10 to 150 qubits. The results indicate promising scalability, with an approximation ratio above $99\%$ and convergence in 6 iterations even in for 150 qubits. Additionally, we run tests on IBM's superconducting hardware for 40, 60 and 80 qubits and obtain solutions compatible with the simulated annealning. The results demonstrate that this ITEMC yields high performance after only a few iterations, making it a practical alternative to conventional resource-intensive algorithms. 

Looking ahead, there are several directions to extend this work. First, exploring additional optimization problems beyond QUBO could demonstrate the versatility of the algorithm in different domains. Another interesting possibility would be to find specific subclasses of QUBO problems for which the method performs best. Second, incorporating machine learning in the gate sorting schemes could enhance the efficiency and robustness of the method. Third, the experimental implementation on current quantum hardware and matrix product state simulations at large system size beyond 3-regular graph density will be an important step in validating our approach for complex problems.

In conclusion, our ITEMC algorithm presents a resource-efficient and scalable approach to approximating solutions to QUBO problems. By requiring relatively few measurements and incorporating adaptive gate sorting, it offers a viable alternative to existing variational approaches. Future research will further clarify its scope and practical impact, potentially opening new pathways in quantum optimization and beyond.

\acknowledgments
We would like to thank Kostas Blekos and Karl Jansen for stimulating discussions. This work is supported with funds from the Ministry of Science, Research and Culture of the State of Brandenburg within the Centre for Quantum Technologies and Applications (CQTA). ADT gratefully acknowledges the support of the
Physikalisch-Technische Bundesanstalt.


\newpage
\appendix
\onecolumngrid
\section{Cost function for optimizing ITEMC parameters}\label{Appendix A}
The parameters $\theta_i, \boldsymbol{\theta}_{ij}$ are optimized to maximize the overlap between the states evolved by the parametric gates $R_y(\theta_i)$, $U_{ij}(\boldsymbol{\theta}_{ij})$ and the corresponding states evolved in imaginary time.\\
The cost functions take the following form \cite{Chai:2024sca}:
\begin{equation}\label{eq: f-detail}
    \begin{aligned}
        &g_{\tau, k}(\theta_{i}) \\
        &=\cos\left(\frac{\theta_{i}}{2}\right) \left(\cosh\left(\tilde{\tau}_{i}\right) - \sinh\left({\tilde{\tau}_{i}}\right)\cdot \langle \sigma_i^z \rangle \right)\\
        &-i\sin\left(\frac{\theta_{i}}{2}\right)\left(\cosh(\tilde{\tau}_{i})\langle  \sigma_i^y \rangle +\sinh\left({\tilde{\tau}_{i}}\right)\cdot \langle \sigma_i^x \rangle \right)\\[10pt]
        &f_{\tau, k}(\theta_{ij, 0}, \theta_{ij, 1}) \\
        &=\cos\left(\frac{\theta_{ij,1}}{2}\right)\cos\left(\frac{\theta_{ij,0}}{2}\right) \left(\cosh\left(\tilde{\tau}_{ij}\right) - \sinh\left({\tilde{\tau}_{ij}}\right)\cdot \langle \sigma_i^z \sigma_j^z \rangle \right)\\
        &-i\cos\left(\frac{\theta_{ij,1}}{2}\right)\sin\left(\frac{\theta_{ij,0}}{2}\right)\left(\cosh\left(\tilde{\tau}_{ij}\right)\langle \sigma_i^y \sigma_j^z \rangle +i \sinh\left({\tilde{\tau}_{ij}}\right)\cdot \langle \sigma_i^x \rangle \right)\\
        &-i\sin\left(\frac{\theta_{ij,1}}{2}\right)\cos\left(\frac{\theta_{ij,0}}{2}\right)\left(\cosh(\tilde{\tau}_{ij})\langle \sigma_i^z \sigma_j^y \rangle +i \sinh\left({\tilde{\tau}_{ij}}\right)\cdot \langle \sigma_j^x \rangle \right)\\
        &-\sin\left(\frac{\theta_{ij,1}}{2}\right)\sin\left(\frac{\theta_{ij,0}}{2}\right)\left(\cosh\left(\tilde{\tau}_{ij}\right)\langle \sigma_i^x \sigma_j^x \rangle + \sinh\left({\tilde{\tau}_{ij}}\right)\cdot \langle \sigma_i^y \sigma_j^y \rangle \right).
    \end{aligned}
\end{equation}
\\
For a given $k$, both functions can be estimated from the expectation values of a few 1-qubit and 2-qubit observables. This means they can be evaluated with a relatively small number of shots. Additionally, these expectation values appear only as coefficients in the optimization problem. In this sense, this is not an ordinary quantum variational approach. Once the coefficients are known, the optimization can be performed using a standard classical optimizer, without the need for quantum resources, and it is relatively straightforward, as it involves only two parameters.

\section{Optimal $\tau$ value}\label{appendix: tau}
\begin{figure}
\centering
\subfigure{
    \includegraphics[width=0.45\textwidth]{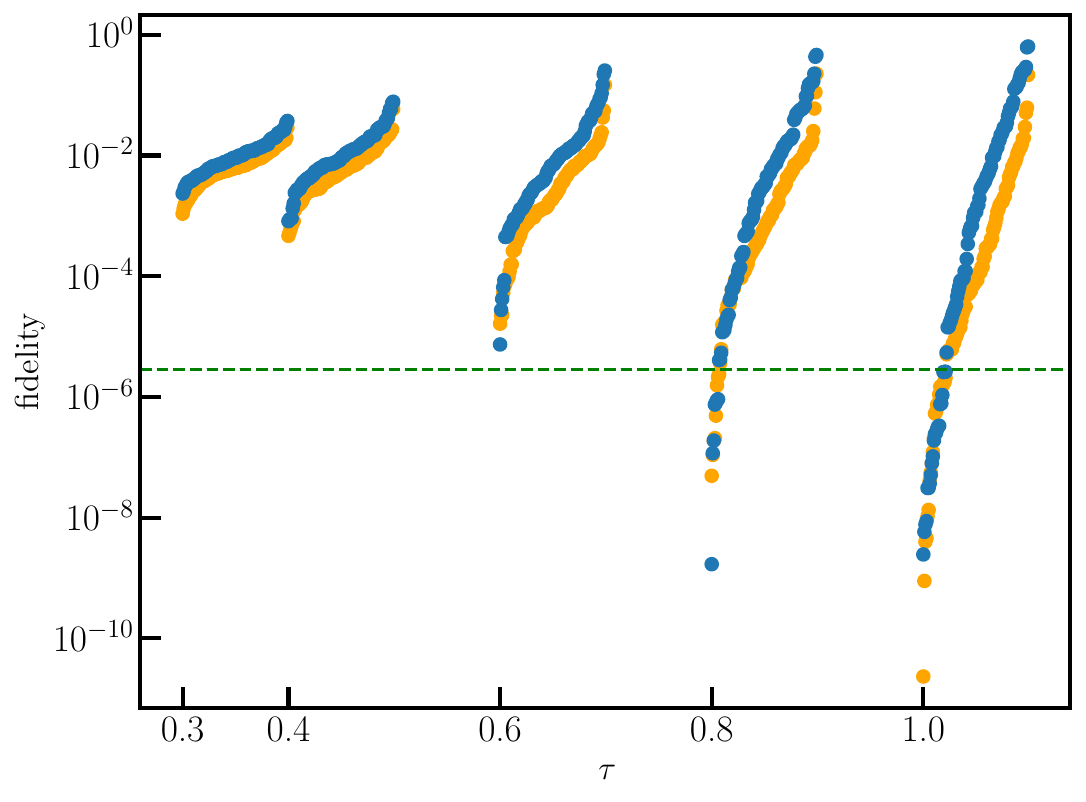}
    \label{fig:20qubits_best3}
}
\subfigure{
    \includegraphics[width=0.45\textwidth]{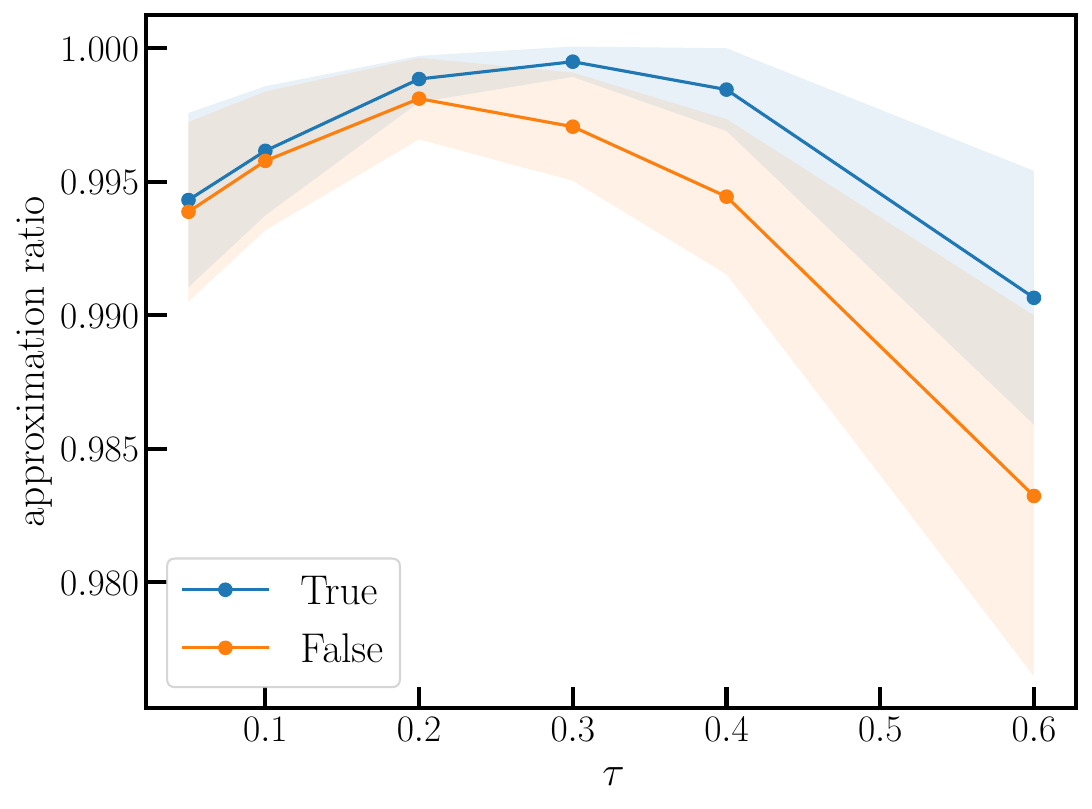}
    \label{fig:sr_ar_N20_tau}
}
\caption{Performance metrics for 20-qubit ITEMC runs avaraged over 100 instances with varying $\tau$. Simulation are ideal with infinite shots. Left panel: Log-scale plot of the fidelity of the ground state for the best 3 solutions. Blue/orange: with/without adaptive sorting. Green line: fidelity of the uniform superposition ($3/2^{20}$). Right panel: Approximation ratio after 5 iterations. $\alpha = 0.01$.}
\label{fig:combined_tau_20q}
\end{figure}

Figure~\ref{fig:combined_tau_20q} displays the performance of the ITEMC after the first iteration for 20 qubits across different values of $\tau$. We observe that a larger $\tau$ can substantially enhance performance for some instances, but it can also degrade performance for worst-case instances. In most of this work we fixed $\tau = 0.3$.

\twocolumngrid
\bibliography{references}

\begin{thebibliography}{47}
\bibitem [{not()}]{note_entropy_scaling}%
  \BibitemOpen
  \href@noop {} {}\bibinfo {note} {The data points in Fig. 7 are insufficient to definitively distinguish between sublinear (e.g., logarithmic) and low-slope linear scaling. A more detailed analysis would be needed to clarify this, but it is beyond the scope of the present work.}\BibitemShut {Stop}%
\end{thebibliography}

\end{document}